\documentclass[sigconf]{acmart} 

\usepackage{acmart-taps}
\usepackage{siunitx}
\usepackage{microtype}
\usepackage{tabularx}
\usepackage{algorithm}
\usepackage{listings}
\usepackage{multirow}
\usepackage{graphicx}
\usepackage{url}
\usepackage{enumitem}
\usepackage{subcaption}
\usepackage{rotating}

\AtBeginDocument{%
  }

\setcopyright{acmlicensed}
\copyrightyear{2024}
\acmYear{2024}
\setcopyright{rightsretained}
\acmConference[HAI '24]{International Conference on Human-Agent
Interaction}{November 24--27, 2024}{Swansea, United Kingdom}
\acmBooktitle{International Conference on Human-Agent Interaction (HAI '24), November 24--27, 2024, Swansea, United Kingdom}
\acmDOI{10.1145/3687272.3688315}
\acmISBN{979-8-4007-1178-7/24/11}

\begin{document}

\title[Impact of Non-Verbal Virtual Agent Behavior on User Engagement in AD]{Exploring the Impact of Non-Verbal Virtual Agent Behavior on User Engagement in Argumentative Dialogues}

\author{Annalena Bea Aicher}
\email{annalena.aicher@uni-a.de}
\orcid{0000-0002-5634-5556}
\affiliation{%
  \institution{Human-Centered Artificial Intelligence, University of Augsburg}
  \streetaddress{Universit{\"a}tsstra{\ss}e 6a}
  \city{Augsburg}
  \country{Germany}
 \postcode{86159}
}
\additionalaffiliation{%
  \institution{Institute of Communications Engineering, Ulm University}
  \streetaddress{Albert-Einstein-Allee 43}
  \city{Ulm}
  \country{Germany}
  \postcode{89081}
}
\additionalaffiliation{%
  \institution{Ubiquitous Computing Systems Laboratory, NAIST}
  \streetaddress{Takayama 8916-5}
  \city{Ikoma}
  \state{Nara}
  \country{Japan}
  \postcode{630-0192}
}

\author{Yuki Matsuda\textsuperscript{$\dagger$}}
\email{yukimat@is.naist.jp}
\orcid{0000-0002-3135-4915}
\affiliation{%
  \institution{Faculty of Environmental, Life, Natural Science and Technology
Okayama University}
  \city{Okayama}
  \country{Japan}
  \postcode{700-0082}
}

\author{Keichii Yasumoto}
\email{yasumoto@is.naist.jp}
\orcid{0000-0003-1579-3237}
\affiliation{%
  \institution{Ubiquitous Computing Systems Laboratory, NAIST}
  \streetaddress{Takayama 8916-5}
 \city{Ikoma}
 \state{Nara}
 \country{Japan}
 \postcode{630-0192}
}

\author{Wolfgang Minker}
\email{wolfgang.minker@uni-ulm.de}
\orcid{0000-0003-4531-0662}
\affiliation{%
  \institution{Institute of Communications Engineering, Ulm University}
  \streetaddress{Albert-Einstein-Allee 43}
 \city{Ulm}
 \country{Germany}
  \postcode{89081}
}

\author{Elisabeth Andr\'e}
\email{elisabeth.andre@uni-a.de}
\orcid{0000-0002-2367-162X}
\affiliation{%
  \institution{Human-Centered Artificial Intelligence, University of Augsburg}
  \streetaddress{Universit{\"a}tsstra{\ss}e 6a}
  \city{Augsburg}
  \country{Germany}
 \postcode{86159}
}

\author{Stefan Ultes}
\email{stefan.ultes@uni-bamberg.de}
\orcid{0000-0003-2667-3126}
\affiliation{%
  \institution{Natural Language Generation and Dialogue Systems, University of Bamberg}
  \streetaddress{Gutenbergstr. 13}
  \city{Bamberg}
  \country{Germany}
  \postcode{96047}
}
\renewcommand{\shortauthors}{Aicher et al.}

\begin{abstract}
Engaging in discussions that involve diverse perspectives and exchanging arguments on a controversial issue is a natural way for humans to form opinions. In this process, the way arguments are presented plays a crucial role in determining how engaged users are, whether the interaction takes place solely among humans or within human-agent teams.\\
This is of great importance as user engagement plays a crucial role in determining the success or failure of cooperative argumentative discussions. One main goal is to maintain the user's motivation to participate in a reflective opinion-building process, even when addressing contradicting viewpoints.
This work investigates how non-verbal agent behavior, specifically co-speech gestures, influences the user's engagement and interest during an ongoing argumentative interaction. The results of a laboratory study conducted with 56 participants demonstrate that the agent's co-speech gestures have a substantial impact on user engagement and interest and the overall perception of the system.\\
Therefore, this research offers valuable insights for the design of future cooperative argumentative virtual agents.
\end{abstract}





\maketitle
\section{Introduction}
Effective and natural communication with humans involves a combination of verbal and non-verbal cues, where gestures and mimics play a crucial role in conveying ideas and concepts beyond words~\citep{liu2021}. Co-speech gestures are a fundamental aspect of non-verbal communication. These spontaneous motions and poses primarily made with the arms and hands (or sometimes other body parts) are produced in rhythm with speech and naturally accompany all spoken language~\citep{wang2021,gesture-frontiers}.

To enhance the effectiveness of virtual and embodied agents in the interaction with humans, it's crucial for them to adopt similar communication strategies~\citep{deichler2023learning}. Humans can integrate information from language and co-speech gesture to derive the message~\citep{kita2023gesture}. As claimed by~\citet{gesture-tedtalks} the study of co-speech gestures and their distinct contributions to (argumentative) discourse could be highly beneficial.

To advance our goal of developing a system that engages users in argumentative discussions while encouraging critical scrutiny of arguments, this paper explores the role of co-speech gestures. The literature on argumentation is fragmented~\citep{miao2022anemer}, and the impact of virtual agents on debates remains unclear~\citep{blount2012avatarinargu}. Addressing this gap, we build on previous research~\citep{iva-aicher}, which found that virtual agents positively influence user engagement\footnote{Defined as ``the quality of user experience that emphasizes the positive aspects of interacting with an online application and the desire to use it longer and repeatedly’’~\citep{lalmas2022measuring}.}, interest\footnote{Defined as ``the activities you enjoy doing and the subjects you like to spend time learning about''~\citep{dict}.}, and perception of the agent. We examine how non-individualized co-speech gestures in human-like virtual agents affect these aspects, as well as user trust and opinion formation in cooperative dialogues.

Our findings confirm that co-speech gestures enhance user engagement, interest, and perception of the virtual agent, even when not tailored to argument content or user responses. Since these gestures do not manipulate users' opinion formation or trust, they effectively strengthen motivation and engagement in argumentative dialogues, promoting critical examination of arguments, the development of well-founded opinions, and longer-lasting interactions.

The paper is structured as follows: Section~\ref{sec:lit} provides a brief overview of related work. Section~\ref{sec:archi} details the architecture of the argumentative dialogue system (ADS). The experiment and study setup are outlined in Section~\ref{sec:userstudy}, with evaluation results in Section~\ref{sec:res}. Section~\ref{sec:dis} discusses these results, followed by a brief conclusion and outlook on future work in Section~\ref{sec:out}. Lastly, Section~\ref{sec:lim} addresses the limitations of this study.

\section{Related Work}\label{sec:lit}
Gesture is one of the most evident forms of nonverbal communication~\cite{wang2021,kendon2004}. Much of the prior work on the nonverbal communication behavior of ECAs has focused on co-speech gestures and their impact on human-agent communication~\cite{kendon2004}. McNeill’s typology~\cite{mcneill}, widely recognized in the field, classifies gestures into four primary categories: (1) deictic gestures, (2) iconic gestures, (3) metaphoric gestures, and (4) beat gestures. Iconics depict concrete concepts by mimicking their size, shape, or contour; metaphorics represent abstract concepts through concrete imagery created by hand and arm movements; deictics are pointing gestures that refer to an entity by extending the index finger, hand, or arm; and beats are biphasic up-down movements of the finger, hand, or arm~\cite{maRelationshipDifferentTypes2022}.

With advances in artificial intelligence, the methods used to generate respective agent behavior, i.e. natural gestures~\cite{liu2021, deichler2023learning,hasegawa2018evaluation, endrass2010} have evolved throughout the years. For instance,~\citet{aamas-gestures-2023} introduced a system that parses raw text in real-time and generates an appropriate emotional and gestural performance which is claimed to also convey personality traits.
When modeling this kind of behavior, the respective impact and influence on the user impression is subjective and depending of various factors.~\citet{pelachaud2010} conducted an experiment with a virtual agent that demonstrates how language generation, gesture rate and a set of movement performance parameters can be varied to increase or decrease the perceived extraversion. Particularly the gesture expressivity of virtual agents has been investigated by~\citet{PELACHAUD2009630}. Moreover,~\citet{ravenet} proposes human gesture characteristics and theoretical frameworks on metaphors and embodied cognition.
Furthermore, \citet{olafsson2020} showed the interaction with the humorous agent led to a significantly greater change in motivation to engage in a healthy behavior (increase in fruit and vegetable consumption) than interacting with the non-humorous agent. Moreover, also in a listening condition the results of~\citet{gratch2007} indicate that  non-verbal communication can create rapport and improve the effectiveness of a virtual agent.
Moreover, several studies~\citep{SINATRA2021106562,dewit2020} indicate that co-speech gestures have a positive impact on the learning process and user engagement in educational settings.
In~\citet{He2022}, they compared gestures produced by a machine-learning model with idle behavior in user perception of a virtual robot presenting classical Roman monuments. They used self-assessment questionnaires to measure human-likeness, animacy, perceived intelligence, and attention. While differences between gesture and idle conditions were minor, the eye gaze tracker showed data-driven gestures attracted more attention. These findings suggest users may respond more strongly to corresponding co-speech gestures in active interactions. Thus, in our study, we focused on a human-like agent engaging in live conversation, aiming to understand users' overall perception, trust, engagement, perceived content, and impact on opinion and interest.

Still the literature focusing on the influence of agents and their nonverbal behavior in argumentative dialog systems is very scarce~\citep{blount2012avatarinargu}. To the best of our knowledge aforementioned findings still lack an analysis of the change in engagement, motivation and perception of a cooperative argumentative dialogue system when a virtual human-like agent uses co-speech gestures compared to a static behavior. Within this paper we aim to close this gap and furthermore analyse whether co-speech gestures are keeping up the user's motivation to maintain the interaction.

\section{ADS Architecture}\label{sec:archi}
In the following, the architecture of our ADS and its components, in particular the underlying dialogue model, argument structure and interface are outlined. 
\subsection{Dialogue Model and Argument Structure}
To be able to combine our ADS with existing argument mining approaches to ensure its flexibility in view of discussed topics, we adhere to the bipolar argument annotation scheme introduced~\citet{stab2014-AAC}\footnote{Due to the generality of the annotation scheme, the system is not confined to the data considered herein. In general, any argument structure that aligns with the applied scheme can be utilized.}. This scheme encompasses argument components (nodes), structured in the form of bipolar argumentation trees. The overall topic represents the root node in the graph. We consider two relationships between these nodes:~\textit{support} or~\textit{attack}. Each component, excluding the root node (which has no relation), has exactly one unique relation to another component. This results in a non-cyclic tree structure, wherein each node, or ``parent'', is supported or attacked by its ``children''. If no children exist, the node is a leaf and marks the end of a branch.
The interaction between the system and the user is separated in turns, consisting of a user action and corresponding natural language answer of the system. The system response is based on the original textual representation of the argument components, which is embedded in moderating utterances.
Table~\ref{tab:moves} shows the possible moves (actions) the user can perform. These enable the user to navigate through the argument tree and enquire more information. Furthermore, the users can state whether they agree or disagree with the given argument. After listening to the minimum of required arguments (20\footnote{To ensure that the interaction lasted long enough and that a sufficient number of arguments were presented.}), the users could exit the conversation.
\begin{table}[h!]
\centering
\caption{Description of possible user actions.}\label{tab:moves}
\begin{tabular}{ll}\toprule
\textbf{Move} & \textbf{Description} \\ \toprule
        \emph{$why_{pro}$} & Request for a pro argument. \\
        \emph{$why_{con}$} & Request for a con argument. \\
        \emph{$suggest$} & Request for an argument (without polarization). \\
        \emph{$level_{up}$} & Returns to the parent node. \\
        \emph{$prefer$} & Agree/Prefer current argument.\\
        \emph{$reject$} & Disagree/Reject current argument.\\
        \emph{$exit$} & Quit the conversation.  \\ 
        \emph{$help$} & Request for help what to do next.\\ \bottomrule
\end{tabular}
\end{table}
In this study a sample debate on the topic \textit{Marriage is an outdated institution} provides a suiting argument structure\footnote{We considered this topic as suitable as topics with a ``more substantial societal need'' are much more likely to cause strong emotions and biases due to their relevance and timeliness. We aimed to minimize these effects to better differentiate between the influences attributed to the topic itself and those associated with the agent's non-verbal behavior.}. It serves as knowledge base for the arguments and is taken from the \textit{Debatabase} of the idebate.org\footnote{\noindent \url{https://idebate.org/debatabase} (last accessed 23\textsuperscript{th} July 2021).
Material reproduced from \url{www.idebate.org} permission of the International Debating Education Association. Copyright \copyright~2005 International Debate Education Association. All Rights Reserved.} website. It consists of a total of 72 argument components (1 \textit{major claim}, 10 \textit{claims} and 61 \textit{premises}) and their corresponding relations and is encoded in an OWL ontology~\cite{bechhofer2009owl} for further use. In each $why_{pro/con}$ move a single argument component is presented to the user. 
To prevent the user from being overwhelmed by the amount of information, the available arguments are presented to the users incrementally on their request. Therefore, the children to a parent node are only presented upon the user's request ($why_{pro}$, $why_{con}$, $suggest$).

\subsection{Interface}\label{sec:interface}
\begin{figure*}[ht!]
  \centering
  \includegraphics[width=0.6\textwidth]{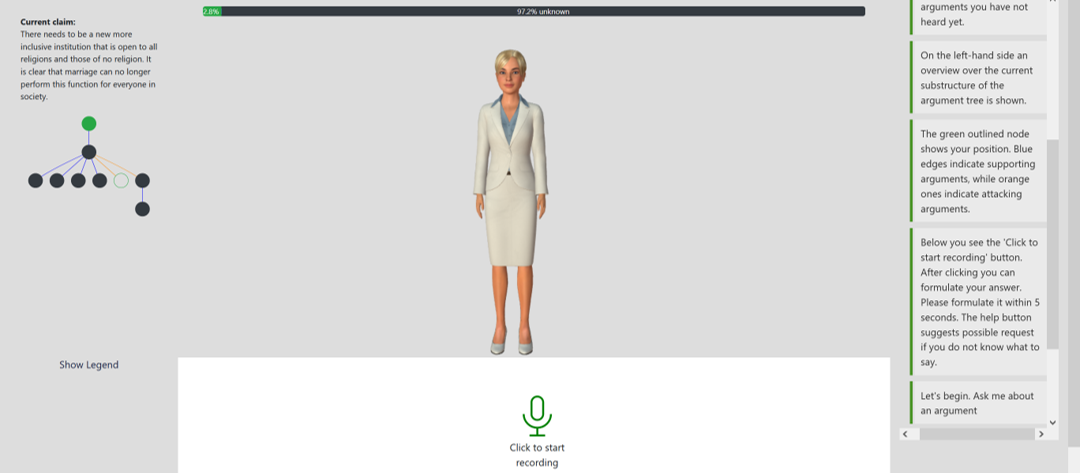}
  \caption{GUI. Above ``Click to start recording'' button the agent ( avatar by Charamel\textsuperscript{TM}) is shown. The dialogue history is shown on the right, the sub-graph of the current branch on the left. The system utterances are marked in green, user responses in blue.}\label{fig:static}
\end{figure*} 
The interface depicted in Figure~\ref{fig:static}  is centered around the Charamel\textsuperscript{TM} avatar~\footnote{\url{https://www.charamel.com/competence/avatare}, licensed under CC BY 4.0 (\url{https://creativecommons.org/licenses/by/4.0}).} which presents the system utterance by lip-sync speech output using the Nuance TTS along with the Amazon Polly voices\footnote{https://docs.aws.amazon.com/polly/latest/dg/voicelist.html}.

\aptLtoX{\begin{table}[h!]
\caption{Exemplary co-speech gestures of the agent (avatar by Charamel\textsuperscript{TM}) in the gesture system.}\label{tab:gestures}
\begin{tabularx}{\columnwidth}{@{}l|Xl@{}}
\toprule
\hspace{0.5cm}\multirow{4}{*}{RANDOM } \hspace{0.1cm}&Predefined co-speech gesture, consisting of mostly beat and some metaphoric gestures~\citet{mcneill}. For example, when expressing ``to get an idea of the whole aspect~[...]'', ``consequently it can be inferred~[...]'', ``it can be deduced~[...]'' etc. & 
      \includegraphics[width=.14\columnwidth, scale=0.4]{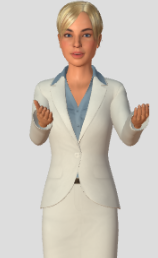}
    \\ \midrule   
\hspace{0.5cm}\multirow{12}{*}{EXPLICIT } \hspace{0.1cm}&Ceictic co-speech gesture pointing to a GUI element at the left bottom, explicitly matching agent utterance, while introducing the respective GUI element. For example, when expressing ``looking at the argument graph~[...]'' etc.&  
      \includegraphics[.14\columnwidth, scale=0.4]{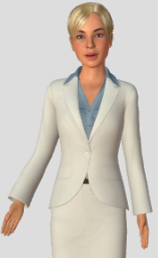}
\hspace{0.5cm} \\ \cmidrule{2-3}
 & Deictic co-speech gesture, pointing to a GUI element at the right middle, explicitly matching agent utterance, while introducing the respective GUI element. For example, when expressing ``as I have mentioned earlier~[...]'' etc.& 
      \includegraphics[.14\columnwidth, scale=0.4]{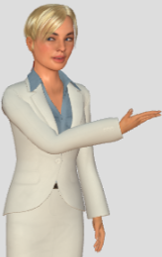}
 \hspace{0.5cm}\\ 
 \bottomrule
\end{tabularx}
\end{table}}{\begin{table}[h!]
\caption{Exemplary co-speech gestures of the agent (avatar by Charamel\textsuperscript{TM}) in the gesture system.}\label{tab:gestures}
\begin{tabularx}{\columnwidth}{@{}l|Xl@{}}
\toprule
\hspace{0.5cm}\multirow{4}{*}{\begin{rotate}{90} RANDOM \end{rotate}} \hspace{0.1cm}&Predefined co-speech gesture, consisting of mostly beat and some metaphoric gestures~\citet{mcneill}. For example, when expressing ``to get an idea of the whole aspect~[...]'', ``consequently it can be inferred~[...]'', ``it can be deduced~[...]'' etc. & 
 \begin{minipage}{.14\columnwidth}
      \includegraphics[scale=0.4]{gesture_1.png}
\end{minipage} \hspace{0.5cm}
    \\ \midrule   
\hspace{0.5cm}\multirow{12}{*}{\begin{rotate}{90} EXPLICIT \end{rotate}} \hspace{0.1cm}&Ceictic co-speech gesture pointing to a GUI element at the left bottom, explicitly matching agent utterance, while introducing the respective GUI element. For example, when expressing ``looking at the argument graph~[...]'' etc.&  
\begin{minipage}{.14\columnwidth}
      \includegraphics[scale=0.4]{gesture_2.png}
\end{minipage} \hspace{0.5cm} \\ \cmidrule{2-3}
 & Deictic co-speech gesture, pointing to a GUI element at the right middle, explicitly matching agent utterance, while introducing the respective GUI element. For example, when expressing ``as I have mentioned earlier~[...]'' etc.& 
\begin{minipage}{.14\columnwidth}
      \includegraphics[scale=0.4]{gesture_3.png}
\end{minipage} \hspace{0.5cm}\\ 
 \bottomrule
\end{tabularx}
\end{table}}

We opted for a full-body representation of the agent (in the middle of the GUI) as it moves across the screen to introduce and highlight various elements of the GUI. This furthermore enabled us to make use of the more than 50 pre-defined conversational motion-captured gestures supplied by \texttt{Vuppetmaster}\footnote{https://www.charamel.com/products/vuppetmaster}. 
As gesture generation and specific animation are not the focus of our work, we use the pre-defined co-speech gestures provided by the~\texttt{Vuppetmaster} without modification. Furthermore, it is important to note that the focus of the study was to examine the influence of an agent using 'suitable' co-speech gestures (movements of arms and hands for explanation, head movements, etc.), which primarily emphasized the verbal introduction to the arguments and their presentation. As the agent is designed to be perceived as a neutral and impartial conversational partner, we chose neutral and friendly facial expressions to avoid biasing the user. As highlighted by~\citet{luo2023effect}, facial expressions, whether positive or negative, have a significantly stronger impact on participants' trust levels and decision-making behaviors compared to interactions lacking expressive facial cues. Therefore, please note that our study intentionally avoided this, and therefore the analysis of explicit facial expressions or emotions was deliberately omitted.

To ensure the suitability of the co-speech gestures for our purpose, they were manually selected from the set of available conversational motion-captured gestures.  In this process, we adhered to criteria defined by two independent experts as ``natural and appropriate for an argumentative discussion with a neutral conversational partner’’. These criteria are as follows: 
\begin{itemize}
\item No large leg movements (\textit{jumping, hopping, dancing, etc.}); lateral steps are allowed.
\item No turning of the upper body and face away from the user at an angle greater than 45 degrees.
\item Movements of the torso are allowed as long as they are not fast, hectic, jerky, or incompatible with the flow of conversation.
\item Hand and arm movements are limited to the area of the torso, not above shoulder height, unless explicitly pointing to an object above.
\item No movements that can be interpreted in the context of emotions (e.g. stomping the foot or waving) or indicate a non-neutral conversational partner (e.g. crossing arms, thumbs up).
\end{itemize}
The co-speech gestures determined according to these selection criteria were not customized to the specific content of the arguments or adapted to individual users. 

To suit the dialogue context, we divided the set of motion-captured gestures into two general groups: ``explicit’’ co-speech gestures, consisting of deictic gestures~\citep{mcneill}, which are only used in specific contexts (e.g., pointing to a GUI element, see Table~\ref{tab:gestures}), and 25 ``random’’ co-speech gestures, consisting of mostly beat and some metaphoric gestures~\citet{mcneill}, which can be used for any utterance of the agent (e.g. arms moving slightly forward without explicitly pointing to anything, see Table~\ref{tab:gestures}). The selection was manually assigned to ensure high relevance, coherence, and consistency.
For instance, a new aspect is introduced with the words ``to get an idea of the whole aspect [...]'' while the agent moves her arms forward and in a circular motion. This metaphoric, non-polarizing co-speech gesture of the agent (see Table~\ref{tab:gestures}: ``random’’) supports the expression of ``whole'' within the dialogue without specifically emphasizing the content of the argument itself. Here, ``random'' does not mean randomly chosen but rather refers to selecting a co-speech gesture from 25 options based on the agent's moderating introduction of an argument. This approach ensures that the agent's gestures are not repetitive\footnote{We ensured that the same speech gesture was not used in the previous 5 turns.} and thus appear natural. However e.g. if the agent clearly refers to an element found in the GUI, this will be emphasized in the corresponding deictic co-speech gesture (see Table~\ref{tab:gestures}: ``explicit’’). An example of this would be the user's statement to revisit a previously presented argument, which the agent indicates by pointing to respective argument in the dialogue history.
The synchronization of co-speech gestures with the utterance was also handled by the~\texttt{Vuppetmaster}. Only one co-speech gesture was selected for each agent turn to avoid overloading the interaction.

The dialogue history is shown on the right side of the screen, marking the system answers with a green and the user answers with a blue line. 
Furthermore, on the left side, the sub-graph of the bipolar argument tree structure (with the displayed claim as root) is shown. The current position (i.e., argument) is displayed with a white node outlined with a green line. Already heard arguments are shown in blue. Nodes shown in grey are still unheard. 
A progress bar at the top of the screen shows the number of arguments that were already discussed and how many are still unknown to the user at each stage of the interaction.

An NLU framework based upon the one introduced by Abro et al.~\cite{abro21-NLU} processes the spoken user utterance. By clicking on ``Click to start recording'' the user starts the recording and can formulate their request within 5 seconds after which the recording automatically stops. 
The spoken input is captured by a browser-based audio recording that is further processed by the \texttt{Python} library \texttt{SpeechRecognition}\footnote{\url{https://pypi.org/project/SpeechRecognition/}, last accessed 17.07.2023} using the Google Speech Recognition API. Its intent classifier uses the BERT Transformer Encoder presented by~\citet{devlin2018bert} and a bidirectional LSTM classifier. The system-specific intents (user moves) are trained with a set of sample utterances of previous user studies.
The response generation is based on the original textual representation of the argument components. The annotated sentences are slightly modified to form a stand-alone utterance serving as a template for the respective system response. Additionally, a list of natural language representations for each system move was defined. During the generation of the utterances, the explicit formulation and introductory phrase are randomly chosen from this list.

In our study setting, which is described in more detail in the next section, the interface for both study groups is completely identical, especially with regard to the system's dialogue strategy and response generation. They differ only in the nonverbal behavior of the agent when the agent is speaking. The listening behavior is also identical.

\section{User Study Setting}\label{sec:userstudy}
\textbf{Recruitment:}~The study was conducted in a university laboratory in a period of three weeks and involved participants with a proficient level of English. The entire process, from the introduction to the completion of pre- and post-questionnaires, was designed to take approximately one hour. Participants were compensated at a rate of \$10 per hour, receiving \$10 for their participation.

\textbf{Participants:} The 56 participants (aged 22–41; 15 female, 41 male) from diverse international backgrounds, including European, Asian, South American and African, were divided into two groups: one group, consisting of 27 participants, interacted with an agent using co-speech gestures (referred to as the ``gesture'' group), while the other group, consisting of 29 participants, interacted with a static agent without any co-speech gestures (referred to as the ``static'' group). It is essential to note that this ``static'' behavior does not imply that the agent is entirely immobile. Instead, it includes subtle movements such as lip synchronization, occasional weight shifting (from one foot to the other), and slight changes in hand and forearm positions. We opted for this rather ``static'' behavior to avoid potential disruptions caused by random movements, particularly since the selected co-speech gestures are context-adaptive (e.g., pointing to specific GUI elements to reference previous dialog history). Expressive random movements might be perceived as unexpected and contextually inappropriate. 

\textbf{Research Questions and Hypotheses:} The primary objective of this study was to address the following research questions: 1) Are co-speech gestures suitable to increase the user engagement and user motivation within an argumentative interaction with a virtual agent? 2) Is there a relation between co-speech gestures and the overall perception of the agent during an ongoing argumentative dialogue?
To investigate these research questions, we formulated the following hypotheses regarding argumentative dialogues to be tested during the study:
\begin{enumerate}
    \item[H1]\makeatletter\def\@currentlabel{H1}\makeatother \label{h:1} co-speech gestures of the virtual agent significantly influence the user engagement.
    \item[H2]\makeatletter\def\@currentlabel{H2}\makeatother \label{h:2} co-speech gestures of the virtual agent significantly influence the user interest.
\end{enumerate}
\textbf{Procedure:}~After a brief introduction to the system (short text and instructions on how to interact), participants were required to answer two control questions. These questions served as a means to verify their understanding of how to interact with the system. Only participants who successfully passed this test were allowed to proceed to a test interaction with the system. In the test interaction, users were able to familiarize themselves with the system until they felt confident enough to initiate ``real'' interaction.
During the real interaction, participants were instructed to listen to at least 20 arguments. Participants were not informed about the different nonverbal communication behavior of the agent.

Before the conversation some demographic data was collected, as well as the user's opinion and interest (5-point Likert scale) in the topic. After the conversation the participants had to rate statements on a 5-point Likert scale (1 - 5 = totally disagree - agree) concerning the interaction. They were taken from a questionnaire according to ITU-T Recommendation P.851\footnote{Such questionnaires can be used to evaluate the quality of speech-based services.}~\citep{itutp851}. Furthermore, we asked the users about their engagement using the questionnaire of~\citet{o2018practical} consisting of 12 items, their perception of the conveyed content by six self-defined items and their trust towards the system using the questionnaire of~\citet{korber2019theoretical}\footnote{This questionnaire was developed of to measure trust in automation.} consisting of 11 items.

\textbf{Collected Data:}~The study collected data through self-assessment questionnaires, participant opinions and interests on the discussion topic, the set of arguments heard, and dialogue history. Data protection regulations and participant anonymity were strictly upheld, and participants could withdraw at any time. The study, featuring a cooperative and non-persuasive design, received Internal Review Board approval following a thorough ethical review and met all internal guidelines.

\textbf{Metrics:}~For the evaluation of the self-assessment questionnaire, we computed the mean ($M$) and standard deviation ($SD$) for each individual item and group\footnote{Please note that, since the scales are ordinal, this information is supplementary and included as a matter of common practice, but it is not suitable for significance estimation. For assessing significance, only the $p$-value and effect size $r$ are considered decisive.}. It's worth noting that, with respect to all items, the assumption of a normal distribution, as assessed by the Shapiro-Wilk Test, had to be rejected ($W=0.770-0.917$, $p < 0.001$). Consequently, to assess the significance of the difference between the means of the two groups, denoted as $\Delta_M$, we applied the non-parametric Mann-Whitney U test~\cite{mcknight2010-MWU} for two independent samples without a specific distribution. To determine the significance of the difference between pre- and post-measurements, we utilized the non-parametric Wilcoxon signed rank test~\cite{wilcoxon07-WSR} for paired samples.All non-exploratory tests were corrected for multiple comparisons. Specifically, Bonferroni-corrected p-values ($p_{corr}$), calculated for a set of four comparisons, were used for all pre- and post-comparisons.

\section{Results}\label{sec:res}
In the following section, we present the result of the previously outlined user study. For all subsequent analyses, significant differences are indicated by a bold $p$-value. On average, participants from both groups interacted with the ADS for an approximate duration of 33 minutes and 41 seconds (SD: 6 minutes and 49 seconds) while listening to around 22 arguments.
The category ``Overall Quality'' (``What is your overall impression of the system?'') employs a distinct 5-point Likert scale (5 = Excellent, 4 = Good, 3 = Fair, 2 = Poor, 1 = Bad). Our analysis shows a statistically significant difference ($p = 0.004$) between the two groups. The gesture system achieved an average rating of 3.74 (SD 0.90), outperforming the static system with a rating of 2.90 (SD 1.01). This difference is considered to be of medium magnitude, as indicated by the effect size $r = 0.385$.-

Due to space constraints, the individual items of the questionnaire are not displayed; however, as it aligns with ITU-T Recommendation P.851\footnote{Such questionnaires can be used to evaluate the quality of speech-based services}~\cite{itutp851}, we refer to this source. It consists of 32 individual items, which describe the user's perception of the agent/system\footnote{Since the agent/system with which the users interacted is named ``BEA'', please note that in all questionnaires, the term "system/agent/application" was replaced with ``BEA'' when referring to this specific agent/system. We have maintained the original phrasing of the questionnaires for better clarity.} and can be grouped into the following aspects: information provided by the system (IPS), communication with the system (COM), system behaviour (SB), dialogue (DI), user's impression of the system (UIS), acceptability (ACC),. Furthermore, we added 7 self-formulated items addressing the aspect of ``argumentation'' (ARG), which are as follows: ``I felt motivated by the system to discuss the topic'' (ARG~1), ``I would rather use this system than read the arguments in an article'' (ARG~2), ``The possible options to respond to the system were sufficient'' (ARG3), ``The arguments the system presented are conclusive" (ARG~4), ``I felt engaged in the conversation with the system.'' (ARG~5), ``The interaction with the system was confusing*''\footnote{Items with * have to be inverted.}(ARG~6), ``I do not like that the arguments are provided incrementally*''\footnotemark[\value{footnote}] (ARG~7).

Regarding the aspects communication with the system (COM) and acceptability (ACC) the individual item analysis between both groups does not reveal any significant differences. With regard to the information provided by the system (IPS) it can be perceived that two single items, addressing if the provided information matched the user's request (IPS 1, $r_{IPS 1}=0.391$) and clarity of information (IPS 2, $r_{IPS 2}=0.413$), have been rated significantly better for the gesture group with a medium effect size. 
Another notable significant difference is observed with regard to the aspect system behavior (SB) in two items. These items pertain to the system's flexibility in response (SB 6) and its response time (SB 7), with effect sizes denoting moderate ($r_{SB 6}=0.311$) and small ($r_{SB 8}=0.263$) effects, respectively. 
Within the aspect dialogue (DI), one item concerning the naturalness of the dialogue (DI 1), reveals a significant difference between the two groups ($r_{\text{DI 1}} = 0.334$). 
Within the aspect dialogue (DI), one item addressing the naturalness of the dialogue (DI 1) stands out with a significantly higher rating in the gesture group, showing a strong effect size of $r_{\text{DI 1}} = 0.603$. 

With respect to the aspect user's impression of the system (UIS), the items addressing the user satisfaction (UIS 1) and the usefulness of the dialogue (UIS 2) receive highly significantly better ratings in the gesture group with moderate effect sizes ($r_{\text{UIS 1}} = 0.489$, $r_{\text{UIS 2}} = 0.467$). Furthermore, the unpleasantness of the dialogue (UIS 4, $r_{\text{UIS 4}} = 0.342$) was rated significantly higher in the static group.

Concerning our self-added aspect argumentation (ARG), we observe highly significant differences in the individual items related to the motivation to discuss the topic ($r_{\text{ARG 1}} = 0.618$), the preference to use the system over reading the arguments in an article ($r_{\text{ARG 2}} = 0.496$), and the ``engagement induced by the system'' ($r_{\text{ARG 5}} = 0.522$).

When the individual items and in case of ones marked with * their inverted counterparts, are aggregated within their respective aspects, no significant differences are observed for COM ($p_{\text{COM}}$=0.423) and ACC ($p_{\text{ACC}}$=0.086). However, significant differences are perceivable in the following aspects: IPS with $p=0.002, r= 0.407$, SB with $p = 0.036, r = 0.280$, DI with $p = 0.010, r = 0.345$, UIS with $p = <0.001, r = 0.600$, and ARG with $p = <0.001, r = 0.560$. 

\begin{table*}[h!]
\caption{Means $M$ and $SD$s of the items of the short user engagement questionnaire~\citet{o2018practical}.}\label{tab:res_engage}
\begin{tabularx}{\textwidth}{lXrrrrrr}
\toprule
&& \multicolumn{2}{c}{Gesture} & \multicolumn{2}{c}{Static} \\ \cmidrule(r){3-4} \cmidrule(l){5-6}
\textbf{Asp.} & \textbf{Question}                                                       & \textbf{$M$} & $SD$ & \textbf{$M$} & $SD$ & $p$ value & effect $r$ \\ \midrule
\multirow{4}{*}{FA} & 1. I lost myself in this experience. & \textbf{3.37} & 1.08 & 2.34 & 1.05 & \textbf{<0.001} & 0.457\\
 & 2. The time I spent using the application just slipped away. & \textbf{3.85} & 0.82 & 2.79 & 1.26 & \textbf{0.002} & 0.422\\
 & 3. I was absorbed in this experience. & \textbf{3.63} & 0.79 & 2.66 & 0.90 &\textbf{<0.001} & 0.496\\ \midrule
\multirow{3}{*}{PU} & I felt frustrated while using the application.* & \textbf{2.37} & 0.84 & 3.03 & 1.09 & \textbf{0.014} & 0.330\\
 & I found this application confusing to use.* & \textbf{2.30} & 0.95 & 3.52 & 1.02 & \textbf{<0.001} & 0.537 \\ 
 & Using this application was taxing.* & \textbf{2.52} & 0.94 & 3.14 & 0.95 & \textbf{0.025} & 0.300 \\ 
 \midrule
\multirow{3}{*}{AE} & The application was attractive.  & \textbf{3.26} & 1.163 & 3.21 & 1.15 & 0.799 & 0.034 \\
& The application was aesthetically appealing. & \textbf{3.56} & 0.70 & 3.10 & 0.77 & \textbf{0.031} & 0.288 \\
& This application appealed to my senses. & \textbf{3.41} & 0.69 & 3.00 & 0.76 & \textbf{0.038} & 0.276 \\\midrule
\multirow{3}{*}{RW} & Using the application was worthwhile. & \textbf{3.59} & 0.75 & 2.93 & 0.96 & \textbf{0.011} & 0.336\\
 & My experience was rewarding. & \textbf{3.56} & 0.79 & 2.66 & 0.96 & \textbf{<0.001} & 0.452\\
 & I felt interested in this experience. & \textbf{4.07} & 0.62 & 3.45 & 1.02 & \textbf{0.020} & 0.312 \\
 \bottomrule
\end{tabularx}
\end{table*}

\begin{table*}[h!]
\caption{Means $M$ and $SD$s of the questionnaire items regarding provided argument content.}\label{tab:res_content}
\begin{tabularx}{\textwidth}{Xrrrrrr}
\toprule
& \multicolumn{2}{c}{Gesture} & \multicolumn{2}{c}{Static} \\ \cmidrule(r){2-3} \cmidrule(l){4-5}
\textbf{Question}                                                       & \textbf{$M$} & $SD$ & \textbf{$M$} & $SD$ & $p$ value & effect $r$ \\ \midrule
C1 I liked the arguments suggested by the system. & \textbf{3.44} & 1.42 & 2.38 & 1.18 & \textbf{0.005} & 0.373 \\
C2 The suggested arguments fitted my preference. & \textbf{3.48} & 1.16 & 2.79 & 1.40 &  0.054 & 0.257 \\
C3 The suggested arguments were well-chosen. & \textbf{3.59} & 0.97 & 2.48 & 1.18 & \textbf{<0.001} & 0.471 \\
C4 The suggested arguments were relevant. & \textbf{3.96} & 0.71 & 2.97 & 1.09 & \textbf{<0.001} & 0.477 \\ 
C5 The system suggested too many bad arguments.* & \textbf{2.04} & 0.98 & 3.21 & 1.50 & \textbf{0.003} & 0.394 \\
C6 I did not like any of the recommended arguments.* & \textbf{1.81} & 0.74 & 2.69 & 1.29 & \textbf{0.009} & 0.347\\ \bottomrule
\end{tabularx}
\end{table*}

\begin{table*}[h!]
\caption{Means and standard deviations of the questionnaire items regarding user trust~\citet{korber2019theoretical}.}\label{tab:res_trust}
\begin{tabularx}{\textwidth}{lXrrrrr}
\toprule
&& \multicolumn{2}{c}{Gesture} & \multicolumn{2}{c}{Static} \\ \cmidrule(r){3-4} \cmidrule(l){5-6}
\textbf{Asp.} & \textbf{Question}                                                       & \textbf{$M$} & $SD$ & \textbf{$M$} & $SD$ & $p$ value \\ \midrule                   
\multirow{4}{*}{UP} & The system state was always clear to me. & \textbf{3.33} & 1.04 & 3.14 & 1.03 & 0.510 \\
 & The system reacts unpredictably.* & \textbf{2.70} & 1.20 & 3.28 & 1.00 & 0.077 \\
 & I was able to understand why things happened. & \textbf{3.93} & 1.18 & 3.28 & 1.16 & 0.053\\
 & It's difficult to identify what the system will do next.* & \textbf{2.85} & 1.17 & 3.38 & 1.15 & 0.109  \\ \midrule
\multirow{2}{*}{F} & I already know similar systems. & \textbf{2.78} & 1.12 & 2.72 & 1.13 & 0.892 \\
 & I have already used similar systems. & \textbf{2.63} & 1.30 & 2.69 & 1.14 &  0.759 \\ \midrule
\multirow{4}{*}{PT} & One should be careful with unfamiliar automated systems.*  & \textbf{3.52} & 0.94 & 3.90 & 0.72 & 0.163\\
& I rather trust a system than I mistrust it.  & \textbf{3.07} & 0.87 & 2.76 & 0.99 & 0.252 \\
& Automated systems generally work well. & \textbf{3.03}  & 0.96 & 2.81 & 0.82 & 0.371 \\\midrule
\multirow{2}{*}{TA} & I trust the system. & \textbf{3.26} & 0.98 & 2.78 & 0.95 & 0.080 \\
 & I can rely on the system & \textbf{3.19} & 0.88 & 2.76 & 0.74 & 0.082 \\\bottomrule
\end{tabularx}
\end{table*}

\begin{table}[h!]
\caption{Means $M$ and $SD$s of the user interest and opinion before (pre) and after (post) the interaction.}\label{tab:interest_opinion}
\begin{tabularx}{1.035\columnwidth}{lrrrr|rrrr}
\toprule
&\multicolumn{2}{c}{Pre interest} & \multicolumn{2}{c}{Post interest} & \multicolumn{2}{c}{Pre opinion} & \multicolumn{2}{c}{Post opinion} \\ \cmidrule(r){2-3} \cmidrule(l){4-5} \cmidrule(l){6-7} \cmidrule(l){8-9}
\textbf{Group} & \textbf{$M$} & $SD$ & \textbf{$M$} & $SD$ & \textbf{$M$} & $SD$ & \textbf{$M$} & $SD$\\ \midrule   
Gesture & 3.04 & 0.81 & \textbf{3.96} & 0.90 & 2.89 & 0.80 & \textbf{3.15} & 0.91 \\
Static & \textbf{2.97} & 0.94 & 2.76 & 0.91 & 2.86 & 0.99 & \textbf{3.07} & 1.03\\
 \bottomrule
\end{tabularx}
\end{table}
Table~\ref{tab:res_engage} displays the results of the short form of the user engagement scale introduced by~\citet{o2018practical}. Interestingly, except for one item (AE 2) all items showed a statistically significant difference with foremost medium to strong effect sizes. Merging these single items (inverted counterparts respectively) into their associated aspects leads to an insignificant difference for AE ($p=0.119$, $r=0.209$) and highly significant differences in FA ($p<0.001$, $r=0.633$), PU ($p<0.001$, $r=0.536$) and RW ($p<0.001$, $r=0.513$) with strong effect sizes.  

In Table~\ref{tab:res_content}, the results related to the conveyed content (arguments) are displayed. Except for item C2 (``The suggested arguments fitted my preference.''), all other items show a foremost significant difference between the groups. This is also reflected in the aggregated individual items (and their inverted counterparts) with a very highly significant difference ($p<0.001$) and a strong effect size $r=0.683$ between the groups.

The results in Table~\ref{tab:res_trust} illustrate the user ratings of the individual items taken from the questionnaire~\cite{korber2019theoretical}, which were examined to assess user trust during the interaction with the ADS. With the exception of item UP 1, the gesture system received higher ratings compared to (for F1: equal to), the static system, although none of these differences reached statistical significance. Nevertheless, a pattern emerges, suggesting that users tend to trust an agent using co-speech gestures more than a static one.
This slight tendency could be attributed to the fact that users perceive agent behavior with co-speech gestures as more natural (see also DI 1). However, relying solely on co-speech gestures is not sufficient to influence, manipulate or enhance user trust.
As shown in Table~\ref{tab:interest_opinion} the difference between the two groups regarding the ``pre-interest'' of the participants (measured on a 5-point Likert scale before the interaction, where 1 represented ``Not at all interested'' and 5 represented ``Very much interested'') is insignificant ($p$ = 0.848). Similarly, the difference regarding the "pre-opinion" (rated on a scale of 1 to 5, where 1 represented "Totally disagree" and 5 represented "Totally agree") is also insignificant ($p$ = 0.862). Whereas in the ``post-interest'' (measured after the interaction), a significant difference with $p_{corr} < 0.001$ ($r$ = 0.558) is notable, the difference in the "post-opinion" between both groups is insignificant ($p$ = 0.764).
For the gesture group a highly significant difference in the user interest before and after the interaction with medium effect size is notable ($p_{corr}$ = <0.001, $r$ = 0.444). In the static group, the difference between pre- and post-interest is insignificant ($p=0.385$), though a decrease is perceivable. Moreover, the difference between pre- and post-opinion is insignificant within each group (Gesture: $p=0.070$, Static: $p=0.083$).

\section{Discussion}\label{sec:dis}
In the following the results of our study (Sec.~\ref{sec:res}), particularly regarding our two hypotheses (Sec.~\ref{sec:userstudy}) are discussed.
With regard to both the individual items and the combined aspect categories of the ITU-T questionnaire~\citep{itutp851}, it becomes evident that the ratings for communication with the system (COM) and acceptability (ACC) did not exhibit significant differences. Regarding the aspect COM the observation aligns with our expectations, as the interaction style with the system did not vary between the two groups. With regard to the aspect ACC, even though no significance is reached, the gesture system is rated higher in both aspects.This suggests that the agent's co-speech gestures are perceived positively, but other factors (see COM 1, COM 2, COM 4) still leave room for improvement.
The significant differences related to the system's flexibility (SB 6), response time (SB 8), and naturalness (DI 1) can be attributed to the fact that, even though there is no objective difference between the systems, a gesticulating agent is more dynamic and conveys the impression of a livelier, more natural conversation.
Consistent with these observations, the respective aggregated aspect categories, SB, DI, UIS, and our self-introduced category ARG also exhibit a significant preference for the gesture system. Hence, it can be inferred that the overall impression of the system, particularly concerning items such as satisfaction (UIS 1), usability (UIS 2) and pleasantness (UIS 4) is enhanced significantly through the use of co-speech gestures. 

It is evident that users experienced a much higher level of engagement in the gesture system, which confirms our first hypothesis~\ref{h:1}. This is investigated in detail through the items presented in Table~\ref{tab:res_engage}. It is apparent that the co-speech gesture system has a notable impact on user engagement, with statistically significant medium effect sizes across all four categories of the user engagement questionnaire, including "focused attention" (FA), "perceived usability" (PU), "aesthetic appeal" (AE), and "reward" (RW)~\citep{o2018practical}. 
This observation is further supported by highly significant differences between the two groups in the individual items ARG 1 (``I felt motivated by the system to discuss the topic.''), ARG 2 (`` I would rather use this system   than read the arguments in an article.'') and ARG 5 (``I felt engaged in the conversation with the system.'') of the ITU-T Recommendation P.851 questionnaire~\citep{itutp851}. 
However, it's worth noting that the ratings for perceived usability (PU) suggest the need for improvement, particularly in addressing errors related to the ASR (Automatic Speech Recognition) and explaining the system's response when the user is not understood correctly (COM 1, COM 2, COM 4).
The results in Table~\ref{tab:res_content} indicate that the co-speech gestures of the agent have a strong influence on the perception of the presented content. As the items address the personal, subjective perception of the provided content, it seems that the objectively samilar presented content (arguments) is significantly better rated due to the corresponding co-speech gestures of the agent.
This is furthermore underpinned by the user ratings concerning the aspect information provided by the system (IPS). Even though the provided content did not objectively differ between the two groups, the subjective impression of the desired information (IPS 1) and clarity of information (IPS 2) is significantly better for the gesture group. 
We can confirm that the opinion-building process of users is not manipulated by subjective impressions. To engage users without influencing their opinions, co-speech gestures were deliberately not tailored to content or emotional expression, avoiding potential bias. The lack of significant differences in user opinions between groups indicates that co-speech gestures effectively engage users in argumentative dialogues with virtual agents while preserving unbiased opinion formation. Additionally, the insignificant difference in user trust between the two groups (Table~\ref{tab:res_trust}) suggests that user trust cannot be solely influenced by co-speech gestures. Therefore, we conclude that it is possible to use co-speech gestures to enhance user engagement and perception without the risk of inducing a bias.
In contrast to the user opinion, there is a statistically significant increase in user interest within the gesture group, aligning with our second hypothesis~\ref{h:2}. While both groups showed no significant difference in interest before the interaction, a significant difference emerged afterward. The gesture group showed a significant increase in interest, whereas the static group did not. These findings suggest that co-speech gestures have a notable influence on user interest and motivation during argumentative dialogue, helping to maintain attention and prevent disengagement.
In conclusion, our findings corroborate our initial hypotheses and demonstrate that co-speech gestures of the virtual agent significantly increased the user interest and engagement compared to a static agent behavior. 

\section{Conclusion and Future Directions}\label{sec:out}
Related literature suggests that the nonverbal behavior of virtual and embodied agents significantly influences the motivation and actions of interacting individuals~\cite{gratch2007}. Given the growing role of social web interactions, it is crucial to understand how agents impact interpersonal communication, especially in argumentation~\cite{blount2012avatarinargu}.
Thus in this work, we investigated the influence of co-speech gestures by a virtual agent on the user's perception, interest, trust, opinion forming and engagement in argumentative dialogues. Therefore a laboratory experiment involving 56 participants was conducted and analysed using self-assessment questionnaires.

Our findings demonstrate that co-speech gestures significantly enhance users' perception, interest, and engagement. Importantly, these gestures positively impact the user's perception of the content without manipulating their opinion formation or trust. This paper thus contributes to understanding how co-speech gestures can enhance user engagement in interactions with cooperative argumentative agents without exerting manipulative effects.
Future research will explore the potential of adapting agent behavior and gestures in response to the presented content to enhance interactions within argumentative dialogue systems. We aim to investigate how natural co-speech gestures and establishing rapport\cite{gratch2007} can sustain the user's motivation to engage with the argumentative dialogue system while fostering an unbiased, well-founded opinion building. Consequently, this study provides important insights for designing future cooperative interfaces involving argumentative virtual agents which can be customized for individual adaptation.

\section{Limitations}\label{sec:lim}
This study has three limitations that future research could address. First, we focused on a proof-of-concept scenario by comparing a ``static'' virtual agent with one using pre-defined, motion-captured gestures from \texttt{Vuppetmaster}, based on carefully selected criteria. As a result, these gestures were not tailored to the specific content of the arguments or to individual user responses. Instead, they were adapted to the agent's statements within the discussion but remained the same for each user. Future research should explore the potential of tailoring gestures to individual and content-specific contexts to enhance their effectiveness.
Second, our study focused solely on co-speech gestures and the speech acts of the virtual agent, without incorporating listening behavior during user turns. To achieve more natural dialogue behavior, future work should also model responsive listening behavior for the agent.
Third, we concentrated on one approach to modeling nonverbal communication. To optimize user engagement and motivation, future work should consider the full spectrum of nonverbal communication, including posture, gaze, facial expressions, emotions, and more, in both the speech and listening behaviors of the virtual agent, while analyzing their respective impacts.

However, it is important to note that that the gesture system displayed a notably higher overall quality when compared to its static counterpart. We contend that this perception of the system's performance can be further improved by tackling these  limitations and tailoring the agent's behavior to better align with the user's nonverbal behavior, expectations and the conveyed content.


\begin{acks}
This work has been funded by the DFG within the project ``BEA - Building Engaging Argumentation'', Grant no. 313723125, as part of the Priority Program ``Robust Argumentation Machines (RATIO)'' (SPP-1999).
\end{acks}

\bibliographystyle{ACM-Reference-Format}
\bibliography{refs}


\appendix



\end{document}